\documentclass[%
aps,
a4paper,
prl,
reprint,
superscriptaddress,
preprintnumbers,
amsmath,amssymb,
floatfix,
]{revtex4-1}

\usepackage{graphicx}
\usepackage{epsfig}
\usepackage{multirow} 
\usepackage{array}
\usepackage{textcomp}
\usepackage{amsmath}
\usepackage{nicefrac}
\usepackage{siunitx}
\usepackage[utf8]{inputenc}

\begin{document}


\title{Real space observation of magnon interaction with driven space-time crystals}

\author{Nick Tr\"{a}ger}
\email{traeger@is.mpg.de}
\affiliation{Max Planck Institute for Intelligent Systems, Stuttgart, Germany}

\author{Pawe\l\ Gruszecki}
\affiliation{Faculty of Physics, Adam Mickiewicz University, Pozna\'n, Poland}

\author{Filip Lisiecki}
\affiliation{Institute of Molecular Physics, Polish Academy of Sciences, Pozna\'n, Poland}

\author{Felix Gro{\ss}}
\affiliation{Max Planck Institute for Intelligent Systems, Stuttgart, Germany}

\author{Johannes F\"{o}rster}
\affiliation{Max Planck Institute for Intelligent Systems, Stuttgart, Germany}

\author{Markus Weigand}
\affiliation{Max Planck Institute for Intelligent Systems, Stuttgart, Germany}
\affiliation{Helmholtz-Zentrum Berlin für Materialien und Energie, Germany}

\author{Hubert~G\l owi\'nski}
\affiliation{Institute of Molecular Physics, Polish Academy of Sciences, Pozna\'n, Poland}

\author{Piotr Ku\'{s}wik}
\affiliation{Institute of Molecular Physics, Polish Academy of Sciences, Pozna\'n, Poland}

\author{Janusz Dubowik}
\affiliation{Institute of Molecular Physics, Polish Academy of Sciences, Pozna\'n, Poland}

\author{Gisela Sch\"{u}tz}
\affiliation{Max Planck Institute for Intelligent Systems, Stuttgart, Germany}

\author{Maciej Krawczyk}
\affiliation{Faculty of Physics, Adam Mickiewicz University, Pozna\'n, Poland}

\author{Joachim Gr\"{a}fe}
\email{graefe@is.mpg.de}
\affiliation{Max Planck Institute for Intelligent Systems, Stuttgart, Germany}

\date{\today}

\begin{abstract}
The concept of Space-Time Crystals (STC), \textit{i.e.} translational symmetry breaking in time and space, was recently proposed and experimentally demonstrated for quantum systems. Here, we transfer this concept to magnons and experimentally demonstrate a driven STC at room temperature. The STC is realized by strong homogeneous micro-wave pumping of a micron-sized permalloy (Py) stripe and is directly imaged by Scanning Transmission X-ray Microscopy (STXM). For a fundamental understanding of the formation of the STC, micromagnetic simulations are carefully adapted to model the experimental findings. Beyond the mere generation of a STC, we observe the formation of a magnonic band structure due to back folding of modes at the STC's Brillouin zone boundaries. We show interactions of magnons with the STC that appear as lattice scattering. This results in the generation of ultra short spin waves down to \SI{100}{\nano\meter} wavelength that cannot be described by classical dispersion relations for linear spin wave excitations. We expect that room temperature STCs will be a useful tool to investigate non-linear wave physics, as they can be easily generated and manipulated to control their spatial and temporal band structure.
\end{abstract}  

\maketitle

Magnons, which are the quanta of spin waves, were intensely discussed in the past decade revealing new fundamental phenomena in magnetism~\cite{RN109,RN3, RN111, RN181,RN219}. Especially intriguing is the analogy to other bosonic quanta, like photons and the corresponding photonics applications. From a fundamental point of view, emerging quantum phenomena are of strong interest. Here, artificial magnonic crystals, \textit{i.e.} systems with periodically modulated magnetic properties, are especially alluring as they allow generating and manipulating spin wave band structures~\cite{RN211, RN219, RN108, RN222}.

Recently, the concept of periodic modulation was extended from space into time, leading to the idea of a time crystal by Wilczek~\cite{RN223}. Watanabe and Oshikawa already noticed that the existence of time crystals should be a logical consequence of Lorentz invariant space-time and long-range order in spatial directions~\cite{RN226}. Indeed, the definition of a time crystalline structure is deduced from ordinary space crystals. The most important criterion for formation of a crystal is the breaking of continuous spatial translation symmetry into a discrete translation symmetry. Equivalently, a time translation symmetry break (TTSB) is essential for time crystals. However, quantum equilibrium states have time-independent observables which forbid TTSB in the ground state~\cite{RN224}. Yet, non-equilibrium states, like periodically driven many-body Floquet systems, can possess a time translation symmetry governed by the external frequency input~\cite{RN231, RN227}. Several experiments confirmed a TTSB in quantum Floquet systems revealing observables with subharmonic responses~\cite{RN236, RN235, RN237}.

Combination of both these symmetry breakings defines a so called Space-Time Crystal that exhibits periodicity in space and time. This was realized by Smits \textit{et al.} as direct observation of space-time crystallinity in a superfluid quantum gas~\cite{RN220}. Additionally, Kreil \textit{et al.} recently proposed a STC at room-temperature in a Bose-Einstein condensate (BEC) of magnons~\cite{RN298}. However, these experiments were limited to quantum systems and only showed the general existence of STCs.

In this work, we unite the fundamental space-time crystal concept within the quantum regime with the world of magnonics and present an exceptional case for nonlinear wave physics in a comparatively large structure. While the existence of STCs has been shown in literature, lattice scattering processes have not yet been observed~\cite{RN220, RN298}. To this end, we generate a driven STC in a Py waveguide and directly image it by time-resolved STXM with x-ray circular dichroism (XMCD) contrast (\SI{20}{\nano\meter} spatial  and \SI{50}{\pico\second} temporal resolution)~\cite{RN111, RN262, RN296}. We use this technique to show the formation of a driven STC and investigate its interaction with magnons at room temperature. Thereby, we observe lattice scattering into higher Brillouin zones and the generation of ultra short spin waves that cannot be explained by conventional dispersion theory. Additionally, we carefully employ micromagnetic simulations to model the experiment to gain a fundamental understanding of the experimental observations.

\begin{figure}
	\includegraphics[width=1.00\columnwidth]{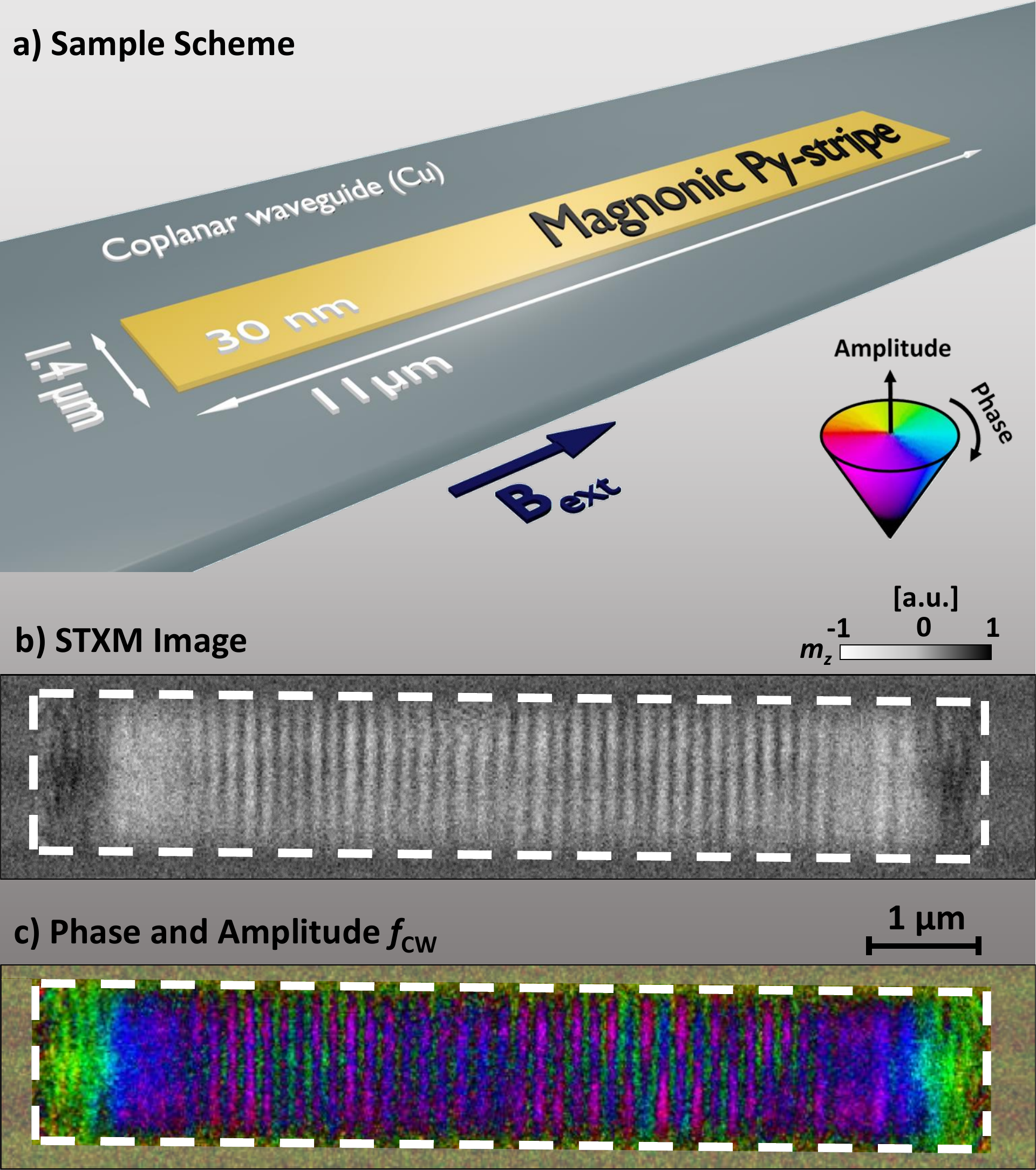}
	\caption{(a) Sketch of the sample with one magnonic Py stripe (yellow) and a coplanar waveguide (grey). (b) Snapshot of a time resolved STXM movie. The greyscale represents the $m_{z}$ component of the magnetization. (c) Phase and amplitude map at $f_\mathrm{CW}$ after FFT in time through each pixel of the STXM movie. The colorcode shows the amplitude and phase information.
	\label{fig1}}
\end{figure}

The magnonic waveguide consists of a \SI{30}{\nano\meter} thick, \SI{1.4}{\micro\meter} wide and \SI{11}{\micro\meter} long Py stripe deposited below a coplanar radio frequency (RF) waveguide. Fig.~\ref{fig1}(a) shows a sketch of the sample with the signal line (shown in dark grey) and the magnonic structure (shown in yellow). A static bias field was applied along the signal line and spin waves in backward volume (BV) were excited by a continous wave (CW) RF field. As the signal line is much larger than the magnonic waveguide, the RF field can be considered as uniform. Thus, spin waves are excited by the combination of oscillations of $\SI{90}{\degree}$ N\'{e}el-type domain walls with the non-uniform dynamic demagnetizing field generated by precessing magnetization at the edges~\cite{RN117, RN197, RN299}. 

\begin{figure}
	\includegraphics[width=\columnwidth]{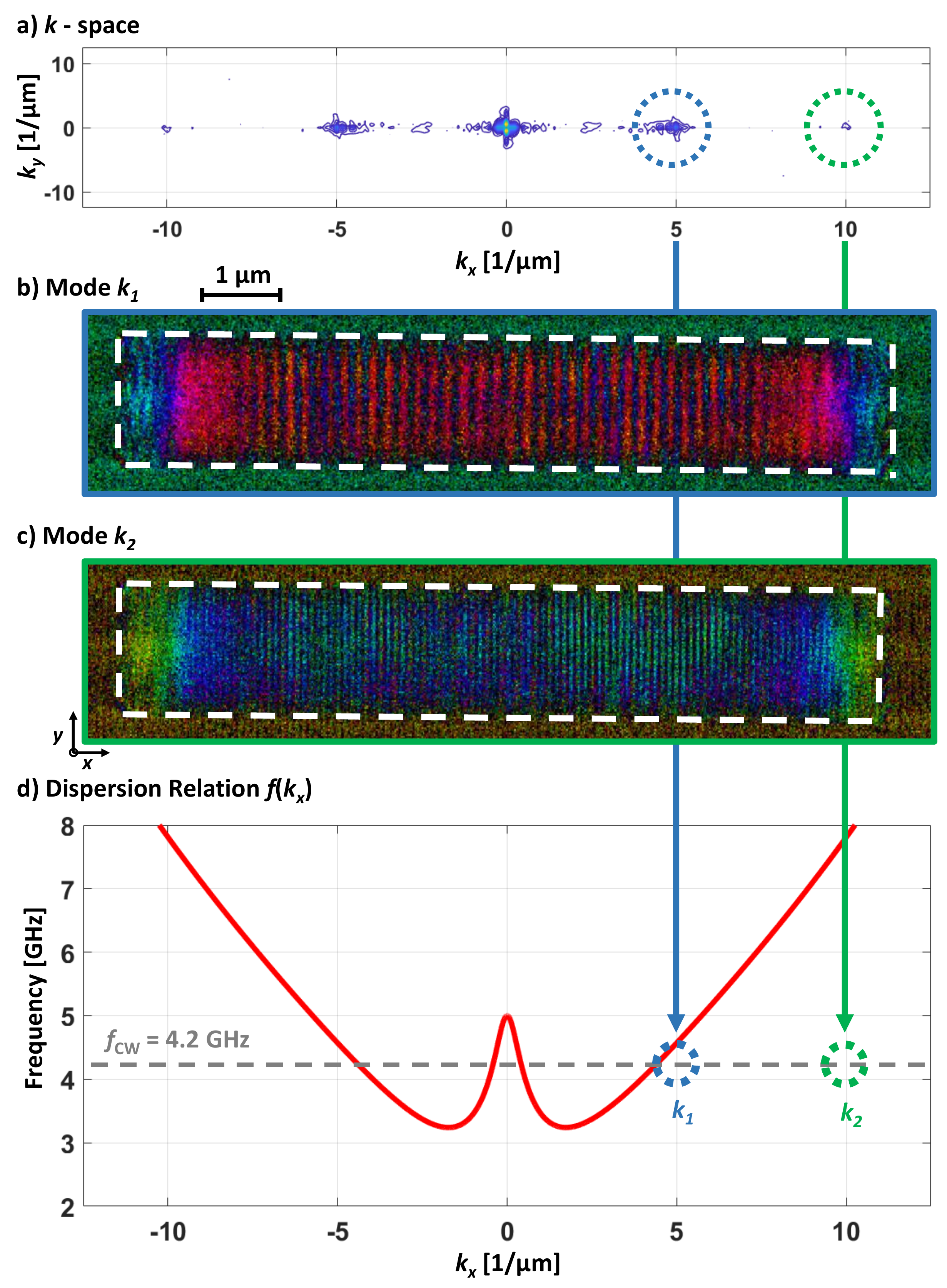}
	\caption{(a) $k$-space retrieved from the spatial FFT of the phase and amplitude map in Fig.~\ref{fig1}(c). Next to the DC peak in the middle, peaks at $k_1 = \SI{5}{\per\micro\meter}$ and $k_2 = \SI{10}{\per\micro\metre}$ can be observed. (b) Phase and amplitude map of $k_{1}$. (c) Phase and amplitude map of $k_{2}$. (d) Dispersion relation $f\left(k_x\right)$ and an excitation frequency $f_\mathrm{CW} = \SI{4.2}{\giga\hertz}$. The calculated dispersion describes mode $k_1$ (blue arrow), but fails to capture $k_2$ (green arrow).\label{fig2}}
\end{figure}

Fig.~\ref{fig1}(b) shows a snapshot of the $m_z$ component of a spin wave from a time resolved STXM movie with CW excitation at $f_\mathrm{CW} = \SI{4.2}{\giga\hertz}$ and an applied field $B_\mathrm{ext} = \SI{8}{\milli\tesla}$. A periodic spin wave pattern is clearly visible. For further analysis, we use a temporal FFT algorithm to access amplitude and phase of the spin wave. This is depicted in Fig.~\ref{fig1}(c) where the amplitude is encoded as brightness and the phase as color.

Subsequently, a spatial FFT allows transition into $k$-space, where the wavevector is $k=\lambda^{-1}$. The $k$-space transformation of the spin waves in Fig.~\ref{fig1}(c) is shown in Fig.~\ref{fig2}(a). Further details on the data evaluation procedure can be found elsewhere~\cite{RN252}.  Next to the DC peak ($k_{x,y}=\SI{0}{\per\micro\metre}$) in Fig.~\ref{fig2}(a), two additional peaks are observed representing distinct spin wave modes. These occur at $k_1 = \SI{5}{\per\micro\meter}$ and $k_2 = \SI{10}{\per\micro\metre}$ which corresponds to wavelengths of $\lambda_1 = \SI{200}{\nano\metre}$ and $\lambda_2 = \SI{100}{\nano\metre}$ respectively. Selective back transformation allows visualization of the real space mode profiles of the individual modes shown in Fig.~\ref{fig2}(b) and Fig.~\ref{fig2}(c), respectively.

Spin wave dispersion theory for infinite films was extended by Guslienko $et\ al.$ and Br\"{a}cher $et\ al.$ to consider lateral confinement in magnonic waveguides~\cite{RN176, RN276}. Fig.~\ref{fig2}(d) shows the theoretical dispersion relation for the confined geometry discussed here. While we find good agreement for the first mode ($k_1 = \SI{5}{\per\micro\metre}$), the second mode ($k_2 = \SI{10}{\per\micro\metre}$) cannot be described using linear theory and cannot be attributed to higher order modes (\textit{c.f.} supplementary material, Fig.~S1). To explain an allowed mode at $k_2 = \SI{10}{\per\micro\metre}$ and its efficient excitation, in the following, we discuss the formation of a magnonic STC and spin wave scattering at the STC Brillouin zone boundary.

\begin{figure*}
	\includegraphics[width=\textwidth]{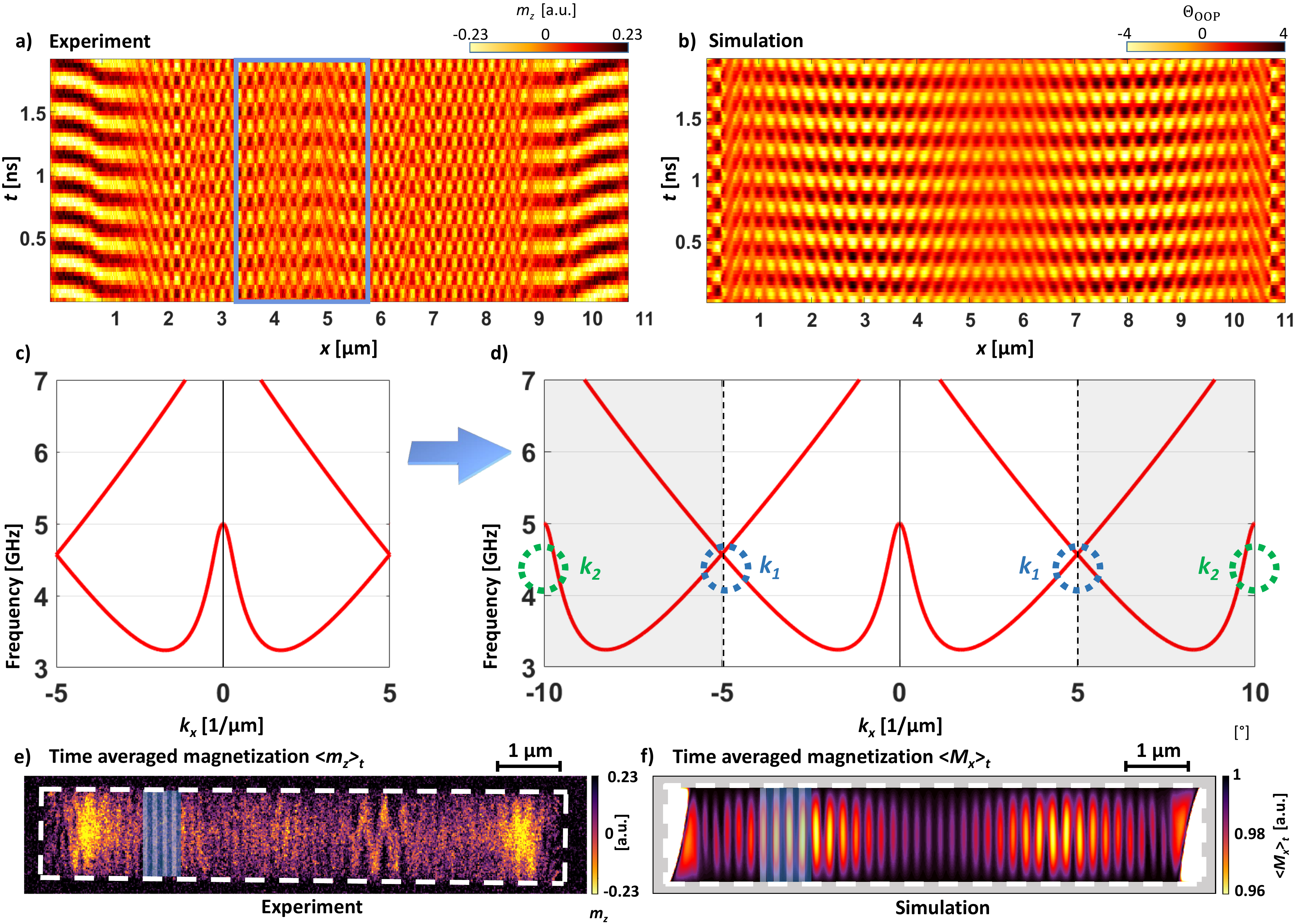}
	\caption{Spatio-temporal evolution of the $m_z$-component from (a) experiment and (b) simulation with good qualitative agreement of a periodic magnetization pattern that is spatially in phase, which indicates condensation of a STC. In simulation, all antinodes are spatially in phase, excluding a simple standing spin-wave that would show a checkerboard-like pattern. In experiment, the STC signature is only observed in the center due to a slight misalignment with respect to the external field (marked area). (c) Dispersion relation $f\left(k_x\right)$ considering this spatial periodicity of \SI{200}{\nano\meter} that defines the Brillouin zone, hence, mode folding occurs at $k_1 = \SI{5}{\per\micro\meter}$. Thereby, a (d) band structure is formed that allows spin wave excitations at high $k$-vectors, like $k_2 = \SI{10}{\per\micro\metre}$ shown in Fig.~\ref{fig2}. (e) Experimental and (f) micromagnetic landscape of the static magnetization of the STC under micro-wave excitation, revealing a spatial periodicity of \SI{200}{\nano\meter} (shown in blue). The high amplitude of the periodic magnetization pattern results in an alternating demagnetization along the bias field direction.\label{fig3}}
\end{figure*}

The continuous application of the RF field leads to the formation of a periodic magnetization pattern and essentially realizes a driven Floquet system with TTSB. While the pattern resembles a standing spin wave at first sight, it cannot be described as such, as all oscillations are spatially in phase as revealed in Fig.~\ref{fig3}(a) and (b) in experiment and simulation, respectively (\textit{cf.} supplemental material, Fig.~S3). Fig.~\ref{fig3}(a) and (b) show the evolution of the out-of-plane magnetization component $m_z$ over time. A simple standing spin wave would result in a checkerboard-like pattern, \textit{i.e.} adjacent antinodes having opposite sign. However, we observe all antinodes to be spatially in phase, excluding a standing spin wave and reinforcing our interpretation as driven STC. While we observe a perfect STC in simulations (Fig.~\ref{fig3}(b)), experimentally, we only detect condensation of an STC in the center of the waveguide (Fig.~\ref{fig3}(a)) as one can see within the marked region. In the experiment slight misalignments of the sample with respect to the external bias field lead to the formation of standing spin wave signatures at the edges. The detection of a STC in the center albeit this misalignment hints at the robustness of STC formation. It is noteworthy, that the periodic magnetization pattern expands into the waveguide at a faster speed than would be expected from the phase velocity of the spin waves and their interference (\textit{cf.} supplemental material, Fig.~S4), further excluding a simple standing spin wave. Hence, the periodic magnetization pattern can be considered as driven STC with the spin wave's wavelength as spatial periodicity ($\lambda_1 = \SI{200}{\nano\metre}$) and a temporal period equal to the driving frequency.

In contrast to quantum systems, like quantum gases or Bose-Einstein condensates, we do not experimentally confirm spontaneous TTSB and subharmonic oscillations. However, the system presented here, still shows spontaneous space symmetry breaking and posses the same periodic modulation in time and space as these STCs. Such a TTSB and accordingly a STC are forbidden in thermodynamic euqilbrium~\cite{RN224}. In contrast, the STC's ground state is a flux equilibirum~\cite{RN298}. Here, this flux equilibrium \nicefrac{dD}{dt}~=~0 of the magnon density $D$ is achieved by continuous creation of magnons by the driving field and damping of magnons. Indeed, our simulations show that non-zero damping is crucial for the formation of a driven STC. 

To gain further insights into the formation of the STC, a micromagnetic simulation of a simplified system was performed (\textit{c.f.} supplementary material, Fig.~S5). Therefor, a narrower (\SI{200}{\nano\meter} wide) but infinitely long Py wave guide is simulated in a spatially uniform RF field at a frequency near the FMR. We find that for high amplitudes of the spatially uniform RF field the precession within the whole sample was uniform. However, a small perturbation, like a minor variations of the RF field on the order of \SI{10}{\nano\meter}, leads to the creation of a periodic magnetization pattern equivalent to the one measured experimentally. Thus, it is possible that such an extremely small perturbation of the RF field causes an injection of large $k$-vectors. In turn, these magnetization patterns lead to a change of the system's properties that are similar to spontaneous translational symmetry breaking.

Above a critical driving power the high amplitude of the periodic magnetization pattern leads to a demagnetizing effect, reducing the magnetization along the in-plane bias field~\cite{RN279}. This is shown experimentally and from micromagnetic simulations in Fig.~\ref{fig3}(e) and Fig.~\ref{fig3}(f), respectively, and elaborated on in the supplemental material Fig.~S2. Through this imprinting of a modulation of the in-plane magnetization, the STC effectively acts as magnonic crystal that forms a band structure for spin waves~\cite{RN108,RN300}. 

\begin{figure}
	\includegraphics[width=\columnwidth]{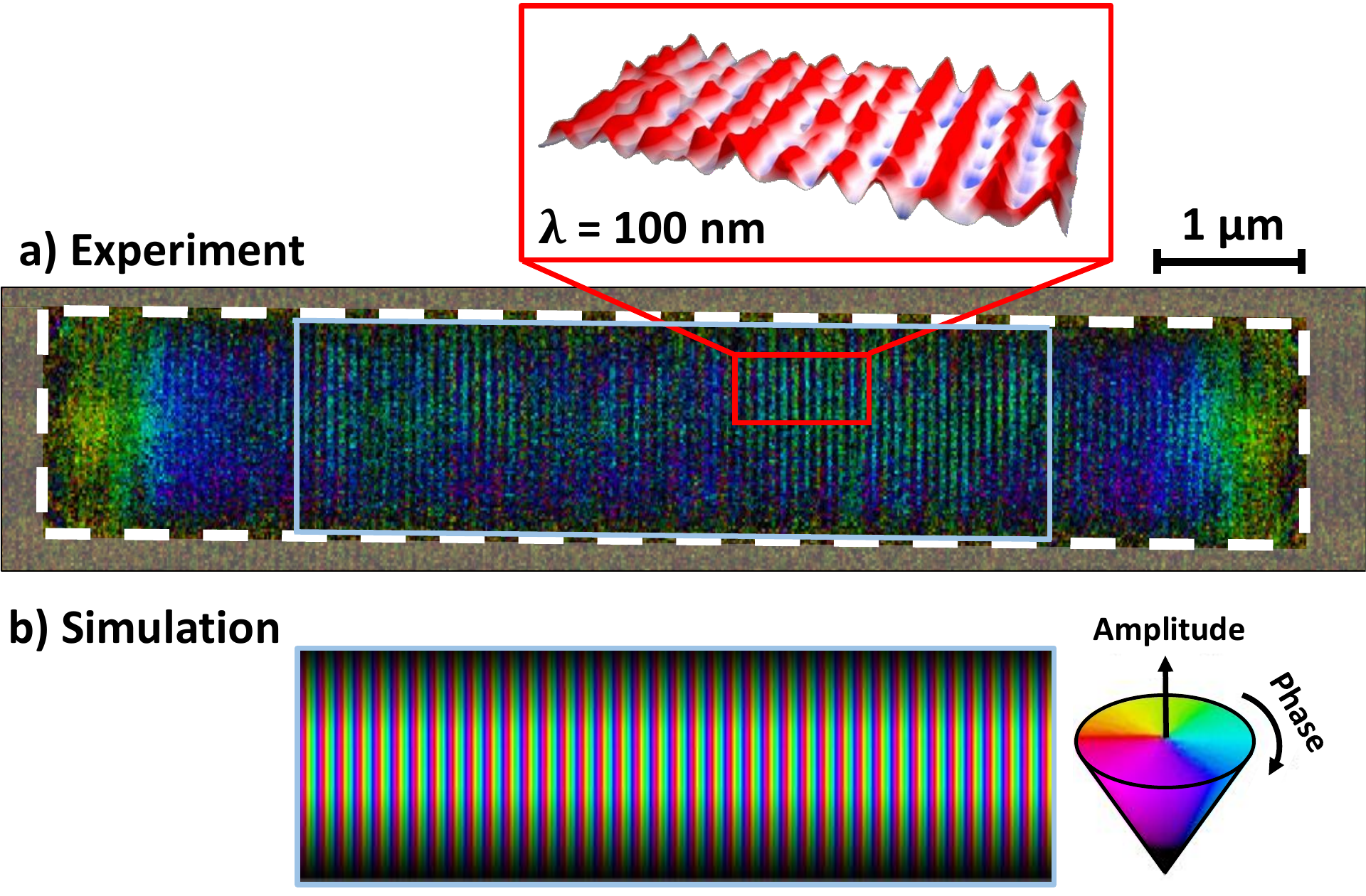}
	\caption{(a) Experimental phase and amplitude map of mode $k_2 = \SI{10}{\per\micro\metre}$, corresponding to a spin wave wavelength of \SI{100}{\nano\meter}. The inset illustrates an enlarged area of the mode profile. (b) Comparison of mode $k_{2}$ from micromagnetic simulations. \label{fig4}}
\end{figure}

The first Brillouin zone of this magnonic STC is shown in Fig.~\ref{fig3}(c). The zone boundary is given by the fundamental spin wave vector $k_1 = \SI{5}{\per\micro\metre}$ at the driving frequency $f_\text{cw} = \SI{4.2}{\giga\hertz}$. However, as the STC extends over the full sample higher Brillouin zones and folding of the modes at zone boundaries also needs to be considered. The resulting extended band structure is shown in Fig.~\ref{fig3}(d) and it becomes evident that higher $k$-modes are also allowed. Fig.~\ref{fig4}(a) and Fig.~\ref{fig4}(b) show the observation such a mode at $k_2 = \SI{10}{\per\micro\metre}$ from experiment and simulation, respectively.

Magnons with $k_2 = \SI{10}{\per\micro\metre}$ are generated by scattering on the periodic magnetization pattern at $k_1$ with the STC. It is noteworthy, that this pattern at $k_1 = \SI{5}{\per\micro\metre}$ form the magnonic crystal where magnons can scatter. Thus, this is a self-scattering process and mode $k_2$ in the second Brillouin zone is only significantly populated at large magnonic densities. However, this scattering process is allowed as both energy and momentum are conserved, because the FMR mode also lies close to $f_\mathrm{cw}$. This leads to efficient RF field absorption in form of uniform in-phase magnetization precession serving as a $k\approx 0$ magnon reservoir. It has been confirmed by micromagnetic simulations that the FMR mode as energy and momentum reservoir is a requirement for efficient STC generation, scattering and population of $k_2$ (\textit{cf.} supplemental material, Fig.~S2). In principle, this can be considered as a four magnon scattering process, yet, one magnon is provided by the STC lattice~\cite{RN278}. Thus, this can also be considered as lattice scattering of a magnon where the loss of STC magnons into the FMR is compensated by the continuous pumping.

In conclusion, we have directly observed a driven Space-Time Crystal and the formation of its magnonic band structure in a Py waveguide experimentally by STXM and by micromagnetic simulations. The room temperature STC is formed by a periodic magnetization pattern that also leads to dynamic demagnetization at non-linear power levels. Furthermore, we have shown the interaction of quasiparticles with such a STC as we observed lattice scattering of spin waves. Folding of the dispersion relation at the STC Brilliouin zone boundaries results in a rich band structure. Thus, lattice scattering processes result in the generation of ultra short spin waves beyond the classical dispersion relation and were observed down to \SI{100}{\nano\meter}.

As driven magnonic STCs can be easily manipulated, this is a unique route to reconfigure magnonic crystals without the need for nanoscale patterning. Furthermore, the reconfigurable STC band structure allows for efficient spin wave generation at ultra short length scales, well below the limits of the classical dispersion relation. However, the observation of interactions with a STC is even more intriguing and is readily accessible by STXM and micromagnetics. In general, we have shown that STCs form band structures at room temperature and that quasi-particles interact with these lattices like in regular crystals. This promises outstanding new opportunities in fundamental research in non-linear wave physics.

\begin{acknowledgments}
The authors would like to thank Michael Bechtel for support during beamtimes. We thank HZB for the allocation of synchrotron beamtime. The research has received partially funding from the Polish National Science Centre project No. 2018/30/Q/ST3/00416. The simulations were partially performed at the Poznan Supercomputing and Networking Center (Grant No 398).
\end{acknowledgments}

\bibliographystyle{ieeetr}
\bibliography{MyEndNoteLibrary}

\end{document}